\documentclass[authoryear,final,5p,times,twocolumn]{elsarticle}

\usepackage{graphicx}
\usepackage{amsmath,amssymb, amsthm}
\usepackage{booktabs}

\journal{Theoretical Population Biology}

\begin{document}
\begin{frontmatter}

\title{Sequential Markov coalescent algorithms for population models with demographic structure}

\author[adr1,adr2]{A.~Eriksson\corref{cor1}}
\ead{anders.eriksson@chalmers.se}

\author[adr1]{B.~Mahjani}

\author[adr1]{B.~Mehlig}

\address[adr1]{Department of Physics, University of Gothenburg, SE-41296 Gothenburg, Sweden.}
\address[adr2]{Department of Marine Ecology, University of Gothenburg, SE-43005 Gothenburg, Sweden.}
\cortext[cor1]{Corresponding author}

\begin{abstract}
We analyse sequential Markov coalescent algorithms for populations
with demographic structure: for a bottleneck model, a population-divergence model,  and for a two-island model with migration.
The sequential Markov coalescent method is an approximation
to the coalescent suggested by McVean and Cardin, and Marjoram and Wall.
Within this algorithm we compute, for two individuals randomly sampled from the population, the correlation between  times to the most
recent common ancestor and the linkage probability corresponding to two different loci with
recombination rate $R$ between them.
We find that
the sequential Markov coalescent method approximates the coalescent well in general in models with demographic structure.
An exception is the case where
 individuals are sampled from populations separated by reduced gene flow.
In this situation, the gene-history correlations may be significantly underestimated. We explain why this is the case.
\end{abstract}

\begin{keyword}
Coalescent \sep Sequential Markov Coalescent \sep Recombination \sep Population Structure
\end{keyword}
\end{frontmatter}

\section{Introduction}

Fast sequencing and genotyping techniques 
have facilitated the 
collection of genome-wide data sets of genetic variation in several organisms, 
including humans \citep{Altshuler:2005,hapmap_group:2007}, mice \citep{lin00:lar}, maize \citep{Yu:2006}, 
and the flowering plant \textit{Arabidopsis thaliana} \citep{Nordborg:2005}.
The empirically observed patterns of genetic variation are shaped by the genetic history of the population in question
which in turn is determined by geographic, historical, and ecological factors. Together with mutations and linkage disequilibrium, these
factors give rise to the observed patterns.

Today the most common theoretical tool for analysing empirical data of genetic variation
is the coalescent algorithm \citep{kin82:gen,griffiths81,hud83:pro}.
In its simplest form, the coalescent algorithm assumes a randomly mixing neutral population of constant size.  The
algorithm has been extended to include intragenic recombination, gene conversion, population structure (both geographic and demographic) 
and temporal variations of the population size,
such as population expansions and bottlenecks \citep[see, e.g., ][for a review]{nor01:coa}.
Last but not least, coalescent theory is now also routinely used in studying the effects of selection \citep{hapmap_group:2007}.

Many sophisticated methods have been developed to analyze empirical genetic data, often based on
sampling gene genealogies by means of coalescent theory under different population models \citep{Marjoram:2006,liang2008saa}.
In order to arrive at accurate inferences with these methods it is essential to sample the genealogies which are most likely to underly the observed data. As a consequence, the analysis typically 
requires very large numbers of samples to be generated \citep{Marjoram:2006}.

Experience shows that the standard implementations of the coalescent algorithm
are very efficient for short 
genomic regions (of lengths of the order of a few hundred kb to a few Mb).
For long genomic sequences, by contrast, little computational efficiency is gained  by skipping over uninteresting generations, 
and substantial computational effort is expended tracking recombination events, their positions, and the many ancestral fragments of each sequence as they repeatedly recombine and coalesce with each other. As a consequence, the computational time increases 
rapidly with the size of the region of interest, and with the sample size.

Several authors have proposed solutions to this problem: \citet{Liang:2007} 
have introduced a method for efficient simulation of many short sequences spread over a long chromosome. Rather than skipping over all generations where the ancestral lines remain unchanged, genealogies are traced generation by generation back through time, and several coalescent and recombination events are allowed to happen simultaneously in each generation. This leads to a significant improvement in performance over standard simulation packages.
However, this method
does not simulate a contiguous locus, but rather isolated loci spread over a large genomic region. Because the work necessary to simulate gene genealogies of large contiguous genomic regions is ultimately bound by the number of genealogical events during the history of the sample, it is not clear how efficient the method of \citet{Liang:2007} would be if it were applied to large contiguous genomic regions.

The sequential Markov coalescent (SMC) method  of \citet{McVean_Cardin05}, and the related method of \citet{Marjoram:2006a}, called SMC' method, 
are approximations to the coalescent algorithm. By construction, these methods result in the same marginal (single-locus) 
gene genealogies as the coalescent, but correlations between different loci are only approximately represented.
The advantage of the SMC and SMC' methods is that they are much more efficient than the standard coalescent algorithm,
in particular when genome-wide gene genealogies must be sampled. 

In the simplest case of a neutral, structureless population with constant population size,
the SMC and SMC' algorithms, 
while approximate, accurately reproduce
the dependence of linkage on the recombination 
rate between two loci on the same chromosome \citep{McVean_Cardin05,Marjoram:2006a}.

In this article we pose the question: what happens when the population size is not constant, or when the population exhibits demographic structure? 
To answer this question, we investigate three common alternatives to a neutral, structureless population with
constant population size:
(a) a population bottleneck model where the population size is taken to be
 piecewise constant, (b) a population-divergence model, and (c) a spatially structured model 
where the population is divided into two sub-populations connected by gene flow. 
For these three models we investigate the performance of the SMC' algorithm.
(Because the SMC and the SMC' methods are so similar in their construction, and the 
SMC' results are found to be significantly closer to the exact results of coalescent theory, we focus on SMC' algorithm in this article.)

The remainder of this article is organised as follows: in Sec.~\ref{sec:background}, 
we review the coalescent algorithm, and briefly describe the SMC and SMC' methods. In Sec. \ref{sec:methods} we define the observables
studied in this article: the probability of linkage and the correlation function of gene histories. We discuss the relation between these
two quantities. Further, we describe our implementation of the SMC' algorithm for the three models of population structure described above.  
In Sec.~\ref{sec:results}, we present our results and analysis. Finally, Sec.~\ref{sec:conclusions} concludes with a brief discussion.

\section{Background}
\label{sec:background}

\subsection{The coalescent}

In this section we briefly describe the coalescent algorithm and its use in interpreting empirical data on genetic variation.
The coalescent algorithm has a wide range of applications, including the inference of population history and structure from empirical observations of the genetic variation of short stretches of a chromosome by means of models of bottlenecks, population expansion \citep{taj89:sta, tajima89b, slatkin_hudson91, sano_etal04},  gene-flow between sub-populations \citep{wakeley96, teshima_tajima03, stumph03}, and complex speciation \citep{Patterson_etal06}.
The coalescent algorithm has also been successfully employed to investigate the effect of short-range variation of the 
recombination rate \citep[recombination hotspots, e.g.][]{wiuf_posada03}, and  of (empirically observed)
long-range changes of the recombination rate along the chromosomes in the human genome \citep{kong_etal02,eriksson_mehlig05}.
Last but not least, a large number of authors have studied the effect of balancing selection \citep{hud88:coa},
 and of directional selection \citep{kap89:hit,krone_neuhauser97,przeworski02,coop_griffiths04,eriksson_etal07} on the genetic variation of a neutral locus on the same chromosome as the selected locus.

Standard implementations of the coalescent method sample the sequence of genealogical events backwards in time \citep{hudson90}.
Usually, one is interested in the gene genealogy of a relatively small sample of a large population. In this case, direct simulations 
of the Wright-Fisher model are inefficient \citep[note, however, the recent advances in direct forward simulations 
which may include more complex models of selection and population dynamics by, e.g.,][]{Hoggart:2007,Carvajal-Rodriguez:2008a}.
Because all genetic variation is assumed to be selectively neutral, 
it is customary to first generate the gene genealogies of all loci and then simulate the mutation process in a second step, given
a particular genealogy.

The genealogical events referred to above include the coalescence of two sequences into a single ancestral lineage, 
recombination with a the sequence, or migration between sub populations (gene flow). 
Coalescent theory builds upon
the assumption that these events 
occur infrequently. 
It is therefore assumed that such events never occur simultaneously, 
and that many generations pass between consecutive events. The number of generations between genealogical events is explicitly computed, 
and is used to skip over generations 
where no genealogical events pertaining to the sample occur. The algorithm proceeds until all loci have coalesced to their most recent common ancestor
in a given sample. 
The resulting genealogy is used to place mutation events according to an appropriate mutation model. 
It is usually assumed that mutations can be modeled by point processes introducing    new genetic variation independently of the ancestral sequence, 
but more complex models have been studied which are argued to more accurately reflect the empirical distribution of nucleotide bases  \citep{Arenas:2007}. 

\begin{figure}[t]
\centerline{\includegraphics[width=252pt,angle=-90]{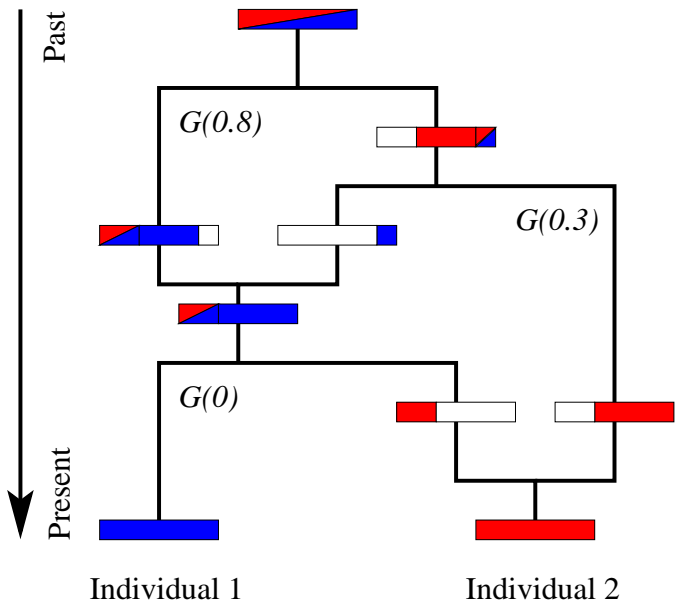}}
\caption{\label{fig:point-wise coalescent}
Illustration of the ancestral recombination graph of a contiguous locus for a sample of two individuals in the coalescent, after \citep{hudson90}. The locus is represented by the interval $[0,1]$, and $G(x)$ denotes the genealogy at position $x$. Two recombination events in the ancestral recombination graph, the more recent one at  $x_2=0.8$ and the older one at $x_1=0.3$, partition the locus into three consecutive regions, $0 \le x < 0.3$, $0.3 \le x < 0.8$, and $0.8 \le x \le 1$, with a different gene genealogy in each region [denoted $G(0)$, $G(0.3)$, and $G(0.8)$, respectively]. $G(0.3)$ is the genealogy resulting from taking the right branch in the first recombination event and the left branch in the second recombination event in the ancestral recombination graph. Similarly, the genealogy $G(0.8)$ corresponds to taking the rightmost branch in both recombination events. 
}
\end{figure}

In the absence of recombination within a contiguous section of a chromosome,
all nucleotides in that section have the same genealogy, and this genealogy is a binary tree. 
Recombination breaks this association between neighbouring nucleotides because the nucleotides to the left of the recombination point have a different parent than the nucleotides to the right, so that the gene genealogies are no longer the same. Recombination events 
are assumed to occur according to a Poisson process along the chromosome. The recombination rate is usually given by $R = 4Nr$ where $N$ is the population size and $r$ is the probability of recombination between a pair of neighbouring nucleotides in a single generation. In this case, the joint gene genealogy of a contiguous segment of a chromosome can be represented as a graph, called the ancestral recombination graph. In Fig.~\ref{fig:point-wise coalescent}, the ancestral recombination graph for a sample of two individuals 
is shown for a stretch of DNA represented by the interval $[0,1]$.  
Two recombination events occurred in the gene genealogy of the sample, at positions $x_1 = 0.3$ and $x_2 = 0.8$ along the stretch
 $[0,1]$. The recombination events separate the gene genealogies into three different groups. If $G(x)$ denotes the gene genealogy of the nucleotides at position $x$ of the chromosome, then $G(x)$ is a piecewise constant function of $x$. It is constant
on the intervals $[0,x_1)$, $[x_1,x_2)$ and $[x_2,1]$. 

This concludes our brief description of coalescent methods.
Tab.~\ref{tab:sim_pack} lists a number of different implementations of the coalescent algorithm.
The first widely spread implementation 
was the {\it ms} program by \citep{hudson02}, and it is still one of the most widely used programs.
In addition to the unstructured coalescent model, this implementation also supports simple models of population expansion and allows for
 sub-populations connected by migration.
The extension of the coalescent framework to include different aspects of evolution is reflected in a large number of simulation programs 
which have emerged over the last few years (Tab.~\ref{tab:sim_pack}).

\begin{table}[t]
\center
\begin{tabular}{ll}
  \toprule
  Program & Authors\\
  \midrule
  \it ms & \citet{hudson02} \\
  \it SimCoal 2.0 & \citet{Laval:2004}\\
  \it SelSim  & \citet{Spencer:2004}\\
  \it CoaSim & \citet{Mailund:2005} \\
  \it CoSi & \citet{Schaffner:2005} \\
  \it SMC & \citet{McVean_Cardin05} \\
  \it SMC' & \citet{Marjoram:2006a} \\
  \it FreGene & \citet{Hoggart:2007} \\
  \it Recodon & \citet{Arenas:2007} \\
  \it Genome & \citet{Liang:2007} \\
  \it msHOT  & \citet{Hellenthal:2007} \\
  \it GenomePop & \citet{Carvajal-Rodriguez:2008a} \\
  \bottomrule
\end{tabular}
\caption{\label{tab:sim_pack}
Simulation packages for the coalescent (the list is not complete). Note that the {\it ms} package of \citet{hudson02} is much older than the publication date indicates.
}
\end{table}

\subsection{The sequential Markov coalescent method for an unstructured population}
\label{sec:SMC}

An important  property of the coalescent process
is that the marginal (single-locus) distribution of gene genealogies is independent of the position of the locus
and thus independent of the recombination model (correlations between genealogies at different loci, by
contrast, depend on the recombination model). 
Based on this observation, \citet{Wiuf_Hein99a,wiu99:anc} introduced an equivalent formulation of the coalescent process where the 
local gene-genealogical tree is followed along the chromosome, rather than tracing the ancestry of all loci back through time, 
as in the coalescent with recombination. 

The procedure for generating the joint gene genealogies of a contiguous stretch of DNA is as follows: 
generate the genealogical tree corresponding to the left-most nucleotide of the stretch for all individuals in the sample.
Let $T$ be the total length of this tree (the sum of the lengths of all branches in the tree), measured in generations.
Generate the location of the next recombination event on the chromosome by moving an exponentially distributed number of nucleotides to the right. The expected value of the exponential distribution is $(rT)^{-1}$.
If the location is beyond the right end of the stretch, the procedure terminates.
If not, consider the genealogy for the nucleotides at the recombination location.  
Pick a point on this genealogy randomly uniformly. At this point (in time) the recombination event is assumed to occur. Trace a line starting 
from this point (the \lq recombination point') back through time, and allow it merge with the ancestral recombination graph $G(x)$ with the standard coalescent rates. This new line is added to the ancestral recombination graph, the genealogy is updated, and correspondingly its total length $T$.
In this way one continues to generate recombination events on the chromosome until the right end of the stretch is reached.

\begin{figure}
\centerline{\includegraphics[width=225pt]{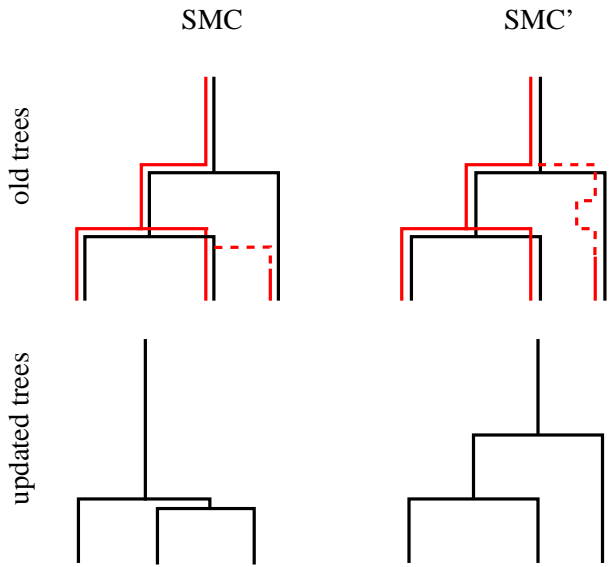}}
\caption{\label{fig:marjoram}
Illustration of the SMC and SMC' approximations 
for a sample of three individuals. The top row shows the line re-attachment process. The given tree is
drawn in black, the resulting tree using red lines. The re-attached line is dashed.
In the SMC method, the line to be re-attached cannot be joined to its old
path, whereas it can do so in the SMC' method. The bottom row shows the resulting gene genealogies after the re-attachment process.
}
\end{figure}

The method of Wiuf and Hein results in the same gene genealogies as the coalescent process. 
But it is also equally computationally demanding for large sample sizes and large genomic regions.
To address this problem, \citet{McVean_Cardin05} sought an approximation to the method by Wiuf and Hein.
As before, gene genealogies 
are
constructed by moving along the chromosome (from left to right). When a recombination event occurs, the gene genealogy is updated.
This update is illustrated in Fig.~\ref{fig:marjoram}a. As usual, the recombination event causes one of the lines in the ancestral recombination graph to detach from the graph.
The part of the line above the point of detachment
(referred to as the recombination point in the preceding paragraph)
is deleted from the ancestral recombination graph (see Fig.~\ref{fig:marjoram}). 
Up to this step, the algorithm does not differ from the one proposed by Wiuf and Hein. 
The difference lies in the following simplification: instead 
of allowing the line to attach to any line of the ancestral recombination graph, 
it is only allowed to attach to lines in the local gene tree
(see Fig.~\ref{fig:marjoram}). 
It is thus no longer necessary to keep track of the full ancestral recombination graph.
Hence, the work necessary to simulate a long contiguous stretch of DNA is significantly reduced compared to the coalescent process.
The drawback of this simplification is that gene histories decorrelate faster as a function of the distance along the chromosome 
as compared with the exact coalescent process.
The reason for this deviation lies in restricting the re-attachment process.

An improvement over the SMC method was suggested by \citet{Marjoram:2006a}. In their
algorithm (SMC' method) the line separated by a recombination event is allowed to attach also to its old path (see Fig.~\ref{fig:marjoram}). 
While retaining the speed and memory efficiency of the SMC method, it was found to be significantly more accurate when compared to the 
exact coalescent process.

\section{Methods}
\label{sec:methods}

\begin{figure}
\centerline{\includegraphics[width=252pt]{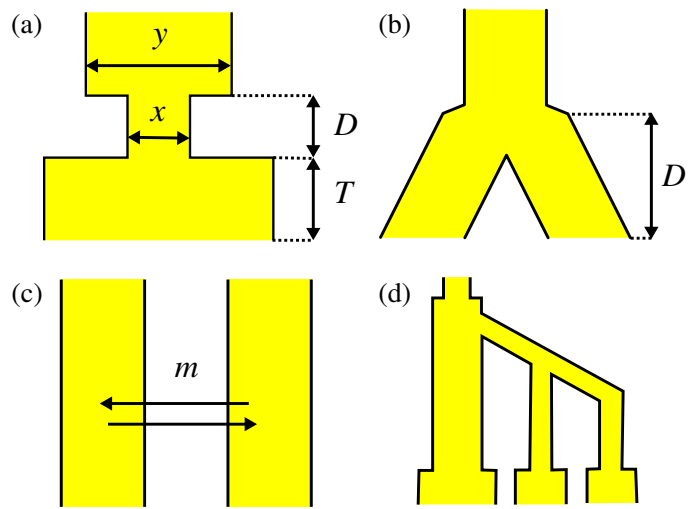}}
\caption{\label{fig:population_models}
Scenarios of population history and structure. 
(a) Population bottleneck model.  
(b) The population-divergence model.
(c) Two-island model. 
(d) A more complex scenario.  }
\end{figure}

The SMC' method in its original form \citep{Marjoram:2006a}  applies to unstructured populations with constant population size.
In this section, we describe how the SMC' method can be extended to models with population structure and changing population size. Figs.~\ref{fig:population_models}a-c show the three models for which we compare the standard coalescent to the SMC' method.  More complex models include combinations of population divergence and expansion, bottlenecks, and migration. As an example of such models, Fig.~\ref{fig:population_models}d depicts the out-of-Africa scenario for  human history used by \citet{Schaffner:2005},
see also \citep{eriksson_mehlig04}. \citet{Schaffner:2005} implement a model of population structure,  bottlenecks, and variable recombination rates along the chromosomes, tuned to match observed of allelic spectrums of different populations (African, Indo-European, Asian, and American).


\subsection{Bottleneck model}

The population bottleneck model is illustrated in Fig.~\ref{fig:population_models}a.
The population was of constant size $Ny$ until $2N(T+D)$ generations ago, when the population decreased to $Nx$ individuals during $2ND$ generations. After the bottleneck, the population quickly expanded to the present population size, $N$ individuals.
It is straightforward to extend the SMC' algorithm to this  model.  First, a gene genealogy for the leftmost locus 
is generated using the standard coalescent method for the model. Second, the position of the next recombination event is
generated as in the SMC' model. Third, the recombining line is re-attached using the same procedure as in the SMC' method,
but where the coalescent rate now depends on time, reflecting the changing population size.

\subsection{Population divergence}

Fig.~\ref{fig:population_models}b shows a model of population divergence.
In this model, the population
was of constant size of $N$ until $2ND$ generations ago, when it diverged into two parts with no gene flow between the two branches,
which each grow to population size $N$. The branches may correspond to different species, or may correspond to sub-populations of the same species separated by e.g. a geographical barrier. If two individuals are sampled from the same branch, the population size is effectively constant. We therefore only consider the case where the individuals are sampled from different branches. In this case, the SMC' algorithm is run separately for the two branches, 
from the present back to the time of the divergence. Some ancestral lines may merge during the divergence. The surviving ancestral lines are pooled in a single population, and the original SMC' algorithm
for constant population size is used to find the gene genealogies.

\subsection{Island model}

We consider the  standard island model of population structure due to \citet{wri31:evo}, see Fig.~\ref{fig:population_models}c.
The population consists of two sub-populations, each with $N$ individuals, connected 
by a gene flow corresponding to $m$ migration events per generation. The scaled
migration rate is  $M = 4Nm$.

The implementation of the SMC' method in this model is 
more complicated than for the models of population bottlenecks and divergence
shown in Fig. \ref{fig:population_models}a and b.
For this reason, the discussion in this subsection
is restricted to sample size two (the algorithms described
above apply to arbitrary sample sizes).

In contrast to the previous models, in the island model it is not sufficient to keep track of the total height of a genealogy in order to update the gene genealogies in the SMC' algorithm. This is because the rate of attachment depends on the number of existing lines present in the island where the  line 
to be attached resides.
Hence, it is necessary to record also the sequence of migration events in the genealogy.

\begin{figure}
\centerline{\includegraphics[width=200pt]{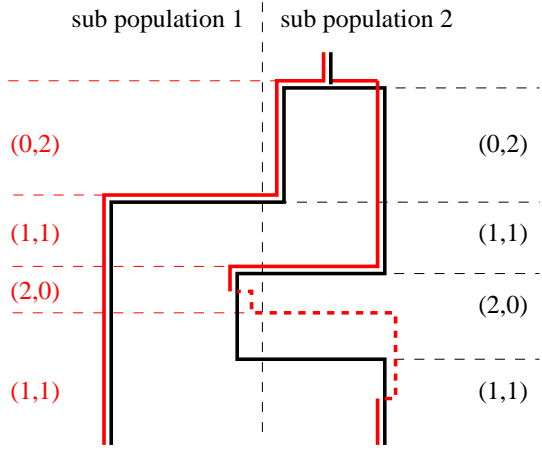}}
\caption{\label{fig.41}
A sample genealogy of genes for the migration model in the $SMC'$ algorithm. 
The black solid lines refer to  the genealogy before recombination occurred, the red lines to the
new genealogy. The re-attached line is drawn as a red dashed line.
Also shown are the sequences of states, both according to the
old and to the new genealogy (in black and red, respectively).
}
\end{figure}

The re-attachment procedure for this population model is implemented as follows.
First, let $n_1$ and $n_2$ be the number of ancestral lines sub populations
1 and 2, respectively ($n_i \in \{0,1,2\}$). Note that the variables $n_i$ are piecewise constant functions of time
(Fig.~\ref{fig.41}).
When a recombination event occurs, the breaking time (referred
to as the recombination point above) is
chosen uniformly randomly on the tree corresponding
to the location of the recombination event. 
The detached line (the dashed line in Fig.~\ref{fig.41}) is
re-attached to the old tree (black lines) applying standard
rates of coalescence, and of migration between sub populations.

\subsection{Gene-history correlations and probability of linkage}

In order to compare the accuracy of the SMC' approximation to the coalescent method, we study the correlation $\rho$ of the times to the most recent common ancestor 
for a sample of two individuals, at two loci separated by recombination rate $R$. This correlation 
underlies empirical measures of genetic variation, such as the variance of the number of segregating sites in a contiguous locus. Hence, if the models differ with respect to the correlation $\rho$, they are likely to differ also with respect to other observables \citep[see, e.g.][for a review of properties of genetic variation in a sample which can be described by the two-locus statistics]{hudson01}.

In the some population models, the correlation $\rho$ is found to be equal to the probability $p_L$ of linkage between the two loci in question. This is the case, for instance, in the constant population-size model and in the population-divergence model (but not in the population bottleneck and two-island models). In some cases (see below) we have been able to compute $p_L$ in closed form in the SMC' approximation. The corresponding results are described in Sec.~\ref{sec:results}, but we first review the known results for the exact coalescent and for the SMC approximation.

For the constant population-size model, \citet{hud83:pro} obtained the by now well-known exact result
\begin{align}\label{eq:p_L_coal_standard}
  p_L(R) &= \rho(R) = \frac{R + 18}{R^2 + 13R + 18}\,.
\end{align}
Within the SMC approximation, the probability of linkage is found to be \citep{McVean_Cardin05}
\begin{align}\label{eq:p_L_smc_mcvean}
  p_L(R) &= \frac{1}{1+R}.
\end{align}
Hence, in the simplest, constant population-size model,
the exact probability of linkage and the one obtained within the SMC approximation
are similar for small and large values of $R$. But they differ for intermediate values of $R$. 

In the population-divergence model, the probability of linkage and the correlation can be calculated exactly using the method described in \citep{eriksson_mehlig04}. One finds:
\begin{multline}
p_L(R) = \rho(R)  =  
 \frac{4 (R^2 + 7 R + 18)  }{(R+2)^2 (R^2 + 13 R + 18)} \,\\+\,
 \frac{ 
  8 (R+6) R\, e^{-D (R+2)/2} + 
  (R+10) R^2\, e^{-D(R+2)}
  }{(R+2)^2 (R^2 + 13 R + 18)}.
\label{eq: 3}
\end{multline}
It should be noted that this result differs from Eq. (14) in \citep{eriksson_mehlig04}. The reason is that here we have conditioned on both individuals being in different sub populations at the present time, whereas the result in \citep{eriksson_mehlig04} was averaged over all four possibilities (both individuals in the same or in different sub populations).

\section{Results}
\label{sec:results}

In this section we describe the results obtained when applying the SMC'
algorithm to the models of population structure described in Fig. \ref{fig:population_models}a-c.

\subsection{Population bottlenecks}

\begin{figure}[t]
\centerline{\includegraphics[width=252pt]{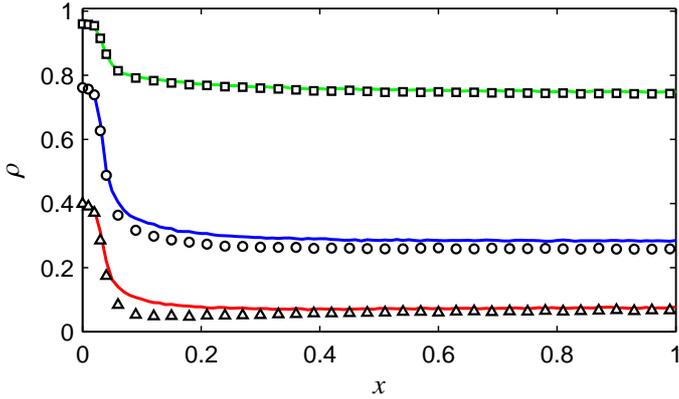}}
\caption{\label{fig:bottleneck_cov}
The correlation $\rho$ of the time to the most recent common ancestor for two loci 
separated by recombination rate $R$ in a sample of two individuals for the population bottleneck model (see Fig.~\ref{fig:population_models}a), for the SMC' method (symbols) and the coalescent (colored lines), as a function of the population size $x$ during the bottleneck, for three recombination rates; $R=0.5$ (green line, squares), $R=3.5$ (blue line, circles), and $R=15$ (red line, triangles). The other parameters are: $T=0.36$, $D=0.18$, and $y=1$. All data points are calculated from simulations of $10^6$ independent realisations.
}
\end{figure}

In the population bottleneck model, the correlation $\rho$ differs from the probability of
linkage $p_L$. Therefore we only study the former.
Fig.~\ref{fig:bottleneck_cov} shows the correlation $\rho$ of the time to the most recent common ancestor for two loci 
separated by recombination rate $R$ in a sample of two individuals for the population bottleneck 
model [see Fig.~\ref{fig:population_models}a]. The correlation
$\rho$ is shown as a function of the population size $x$ during the bottleneck. The symbols correspond to the SMC' 
approximation, and the coloured lines to the coalescent algorithm. Results for three different recombination rates
are shown: $R=0.5$ (green line, squares), $R=3.5$ (blue line, circles), and $R=15$ (red line, triangles). The other parameters are: $T=0.36$, $D=0.18$, and $y=1$. 

Fig. ~\ref{fig:bottleneck_cov} makes it clear that in the bottleneck model, the SMC' approximation
give very similar results to the coalescent. The largest deviations occur for relatively small widths of the bottleneck ($x \approx 0.1$) and for intermediate values of the recombination rate $R$. 

\subsection{Population divergence model}
\label{sec:Population_divergence_model}

\begin{figure}[t]
\centerline{\includegraphics[width=252pt]{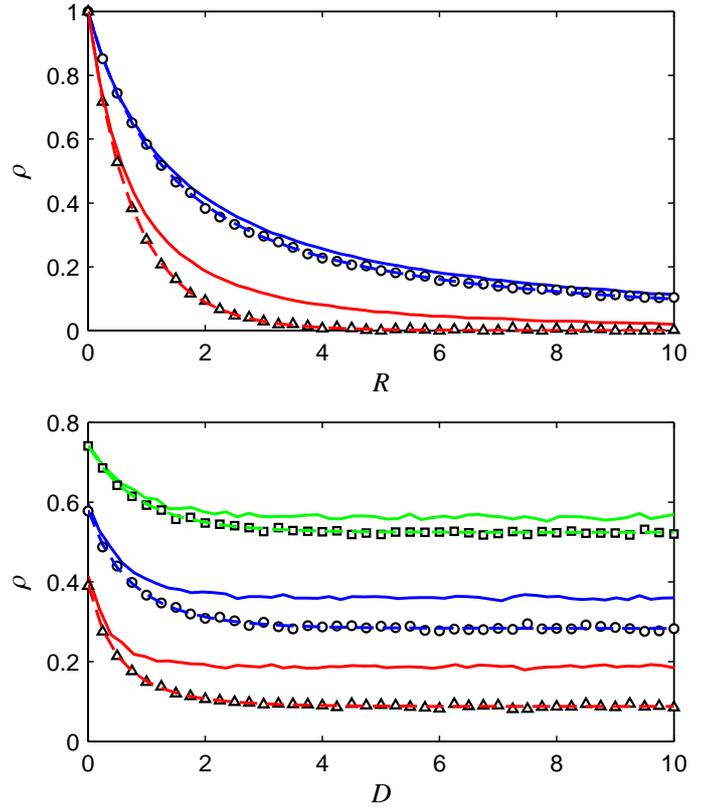}}
\caption{\label{fig:pop_split_corr}
Top: the correlation $\rho$ of the time to the most recent common ancestor for two loci 
separated by recombination rate $R$ for a sample of two individuals
in the population-divergence model (Fig.~\ref{fig:population_models}b); SMC' approximation (symbols) and 
 coalescent (lines), both as a function of the recombination rate $R$. 
All data points are calculated from simulations of $10^{\,5}$ independent realisations.
The dashed lines represent the probability of linkage in the SMC' model, Eq.~(\ref{eq:p_L_split}).
Two choices of the time $D$ since the divergence are 
shown: $D = 0$ (no divergence, blue lines and circles), and $D=10$ (long divergence, red lines and triangles). 
Bottom: The correlation $\rho$ of the time to the most recent common ancestor of the two loci in the population-divergence model, for the SMC' method (symbols) and the coalescent (lines), as a function of the duration $D$ of the divergence. 
All data points are calculated from simulations of $10^5$ independent realisations.
Again, the dashed lines represent the probability of linkage in the SMC' model.
From top to bottom: $R=0.5$, $R=1$, and $R=2$.
}
\end{figure}

Fig.~\ref{fig:pop_split_corr} shows the 
correlation $\rho$ of the time to the most recent common ancestor for two loci 
separated by recombination rate $R$ for a sample of two individuals 
in the population-divergence model. Shown are the SMC' approximation (symbols) and for the standard coalescent (solid lines), 
as a function of the recombination rate $R$ between the loci.

The upper panel shows how the correlation decrease as a function of $R$, for two values of the duration $D$ of the divergence. For $D = 0$, corresponding to the standard constant population size, we confirm that the  SMC' approximation gives a correlation $\rho(R)$ which is very similar to that of the coalescent (blue solid line and circles). For $D = 10$ (red solid line and triangles), however, SMC' approxoimation differs significantly from the
coalescent: the correlations decrease more rapidly in the SMC' approximation than in the coalescent. While
the correlations in the coalescent decrease as a power law, 
they decrease exponentially in the SMC' approximation (as will be shown analytically below) when $D > 0$. 
The largest differences occur for intermediate values of the recombination rate $R$.

A similar pattern can be seen in the lower panel, where the correlation $\rho$ is shown as a function of $D$ for different values of the recombination rate. The correlation is a decreasing function of $D$, but for large values of $D$ the decrease is very slow (we show below that the correlations approach a constant level for large values of $D$). For small values of $D$ the SMC' approximation yields
 correlations $\rho$ which are close to those of the coalescent.  The accuracy of the SMC' approximation declines with increasing values of $D$, 
but for large values of $D$ the error approaches a constant.

It is very difficult to obtain an analytical expression for the correlation $\rho$ in the SMC' approximation.
However, a comparison in Fig.~\ref{fig:pop_split_corr} of the probability of linkage $p_L$ to the correlation $\rho$ 
in the SMC' approximation shows that, for the parameter values investigated, both appear to be identical (to within stochastic fluctuations). 

In the following we show how to analytically calculate $p_L$.
We first derive a general relation between the single-locus distribution function of the time $\tau$ to the most recent common ancestor, $f(\tau)$, and the probability of linkage, $p_L$. Let $\phi(\tau)$ be the probability that a line re-attaches to the same branch when started from a point uniformly on $[0,\tau]$. We refer to the probability $\phi(\tau)$ as the \lq loop probability'.
Because
the number of recombination events separating two loci is Poisson distributed, and since each recombination event leaves the genealogy intact with probability $\phi(\tau)$, we find that the probability of linkage is
\begin{align}
  p_L(R) &= \int_0^\infty d\tau \, f(\tau) \sum_{k=0}^\infty [\phi(\tau)]^k \frac{(R\tau)^k}{k!} e^{-R\tau}\nonumber\\
         &= \int_0^\infty d\tau \, f(\tau) \, e^{-[1-\phi(\tau)] R \tau}. \label{eq:p_L_gen}
\end{align}
Thus, in order to obtain the probability of linkage for a given population model, one needs to calculate the loop probability. 

In Appendix~\ref{sec:prob_linkage} we summarise this calculation for the population divergence model 
(and, as a special case, for a freely mixing population of constant size). The result is:
\begin{align}\label{eq:p_L_split}
  p_L &= \frac{e^{-R/2}}{2} \int_0^1 ds\, s^{(R-2)/4} \exp\left[-\frac{R(1-2e^{-D})(1+s)}{4}\right]\,.
\end{align}
This result can be expressed in closed form in terms of the incomplete Gamma function (see Appendix~\ref{sec:prob_linkage} for details). 
In Appendix~\ref{sec:pL_upper_bound} we derive a bound for this probability:
\begin{align}\label{eq:p_L_split_approx}
  p_L &\le \frac{8 e^{-(3-2 e^{-D}) R/4}}{R^2+8 R+12} + \frac{2 e^{-(1-e^{-D}) R}}{R+6}  \,.
\end{align}
Eq. (\ref{eq:p_L_split_approx})  shows that the linkage probability falls off exponentially with increasing $R$ 
for all values of $D>0$.
Now consider the exact probability of linkage in the coalescent algorithm, Eq. (\ref{eq: 3}). According to this expression, $p_L$ falls off at a rate $\sim 4/R^2$ when $RD \gg 1$, and not exponentially as in the SMC' approximation.

Note, however, that 
when $D = 0$ (corresponding to a freely mixing population of constant size), the bound (\ref{eq:p_L_split_approx}) for $p_L$
in the SMC' approximation does not decrease exponentially but as $2/(R+6)$ for large values of $R$. 
A lower bound for this case, $p_L \ge 1/(1+R)$, is derived in Appendix~\ref{sec:pL_lower_bound} and confirms that for $D = 0$ the probability of linkage decreases inversely proportional to $R$. 


In summary, for the population divergence model we have found the following. For sufficiently large duration of the divergence
(for a large value of $D$), there may be significant differences in the correlation $\rho$ between the coalescent and the SMC' approximation. 
The main reason for this discrepancy is that in the  coalescent, the ancestral lines may merge and break several times during the divergence, but in the SMC' approximation the probability this to happen is much lower. 
In addition, we have established numerically that 
in the SMC' approximation, the correlation $\rho$ and the probability of linkage appear to be identical 
(this is known to strictly hold in the coalescent, c.f. the discussion in the methods section).
Last but not least, our analytical results within SMC' approximation show that the probability of linkage decreases 
exponentially with increasing recombination rate, in contrast the exact coalescent where the probability of linkage decreases as a power law in $R$.

\subsection{Two-island model}
\label{sec:twois}

\begin{figure}
\centerline{\includegraphics[width=252pt]{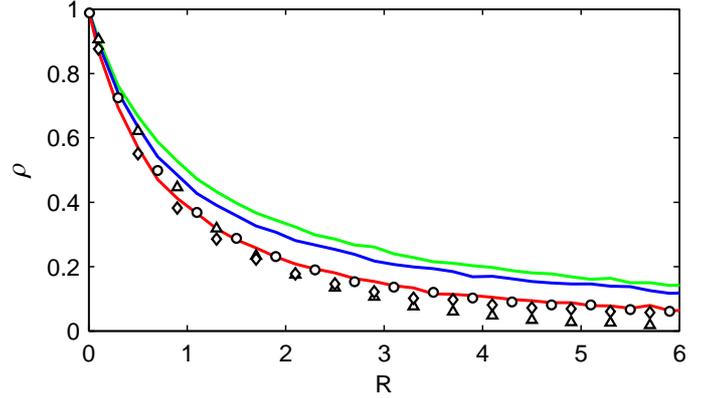}}

\caption{\label{fig:island_corr}
Correlation
$\rho$ of the time to the most recent common ancestor for two loci
separated by recombination rate $R$ in a sample of two individuals 
for the two-island model (see Fig.~\ref{fig:population_models}c), as function of $R$ for $M = 0.01$ (green line and triangles), $M=0.2$ (blue line and circles), and $M=10$ (red line and diamonds). The lines corresponds to from simulations of the coalescent, and the symbols to simulations of the SMC' model, from $10^4$ independent realisations.
}
\end{figure}

In Fig.~\ref{fig:island_corr} we show the correlation $\rho$ of the time to the most recent common ancestor for 
two loci separated by recombination rate $R$ in a sample of two individuals for 
the two-island model (Fig. \ref{fig:population_models}c), as function of $R$ for three different migration rates: $M = 0.01$ (green line and triangles), $M=0.2$ (blue line and circles), and $M=10$ (red line and diamonds). The results are based on $10^4$ simulations of the standard two-island coalescent with migration, and the two-island extension of the SMC' method described in Section~\ref{sec:twois}.

When the migration rate is large, the SMC' approximation gives results very similar to those of the coalescent. 
This is expected since in this case the population behaves as a single panmictic unit. For intermediate and small migration rates, however, differences
are observed. These differences increase
 with decreasing migration rates.

\section{Discussion}
\label{sec:conclusions}

In this article we have investigated the accuracy of the SMC' approximation to the coalescent
for the population models depicted in Fig.~\ref{fig:population_models}a-c.  This is an important question
since many populations show deviations in their allelic distributions consistent with historic population expansions, bottlenecks, and structure from geographic separation or preferential mating. 
To determine the accuracy we have computed the correlation of the time to the most recent common ancestor for two loci
in a sample of  two individuals in the SMC' approximation. We have compared these results to the corresponding exact coalescent results.
In this section we briefly discuss the results for the three different models, and put them in a wider context.

In the population-bottleneck model (Fig.~\ref{fig:population_models}a) we have found that the SMC' approximation works well.
This result can be understood more generally as follows. Consider a general 
model of time-dependent population-size variations, where $N(t)$ is the population size as a function of time. The 
coalescent for this model can be mapped to the constant population-size model by introducing a (possibly nonlinear) transformation of the time $\tau$ to a 
\lq stretched' time 
\begin{equation}
\tau' = \int_0^\tau {\rm d}t\, N(0)/N(t)\,.
\end{equation}
Because of the construction of the SMC' algorithm, 
the coalescent and the corresponding SMC' approximation are transformed in the same way by this transformation. Hence, it may be  expected that the  SMC' 
approximation works well  for arbitrary models of population expansions and bottlenecks. 
We have verified this expectation for the particular case of the bottleneck model depicted in Fig.~\ref{fig:population_models}a.

In the population-divergence model (Fig.~\ref{fig:population_models}b) the situation is different. 
We have calculated the linkage probability $p_L$ (which is equal to $\rho$ in this model) both exactly and within the SMC' approximation of the coalescent. 
Our analytical results show that when the expected number of recombination events during the divergence becomes large, 
the accuracy of the SMC' approximation deteriorates: the exact correlation decreases as a power law as a function
of $R$, whereas in the SMC' approximation it decreases exponentially. 

In order to better understand this difference, let us compare how the divergence affects the gene genealogies in the coalescent to how it affects gene genealogies in the SMC'. In the coalescent, recombination can cause the genetic material to spread out over different ancestral lines during the divergence. The probability of linkage depends on whether the genetical material is spread over two, three or four lines at the onset of the divergence. In each case, however, the coalescent results in a probability of linkage which decreases as a power law with increasing $R$.

In the SMC' approximation, by contrast, the probability of linkage is a function of the loop probability. As explained in Section~\ref{sec:Population_divergence_model}, the cause of the exponential decrease of the probability of linkage in the SMC' approximation can be traced back to how the loop probability depends on the duration of the divergence. 
Hence, the different behaviours of the coalescent and the SMC' approximation can be attributed to the lack of memory of past gene genealogies within the
SMC' approximation. 

Note also that if there is a severe population bottleneck during the divergence, we expect that the differences between the coalescent and the SMC' will be smaller, since the bottleneck increases the chances that the genes will be linked right before the divergence. This means that as far as the correlation 
of times to the most recent common ancestor is concerned, bottlenecks right after the divergence have the same effect as decreasing the time $D$ to the divergence.

The third model we  have considered is the two-island model with migration (Fig.~\ref{fig:population_models}c). 
The extension of the SMC' approximation to this model turned out to be more complicated than in the other two models, since it is necessary
to keep track of not only the time to the most recent common ancestor in the genealogy, but also of the sequence of  migration events. Simulations of the exact
coalescent and the SMC' approximation showed that when the migration rate is large, the SMC' approximation is accurate. 
For small and intermediate values of the migration rate, by contrast,  significant differences were found. 
The magnitude of the relative deviations between the exact result and the SMC' approximation is similar to those in the 
population divergence model. Numerical analysis shows that when the recombination rate $R$ between the two loci is sufficiently large,
 $\rho$ decreases exponentially with increasing $R$ in the SMC', whereas in the coalescent it is approximately inversely proportional to $R$ in this regime.

It follows from our results that in a more complex scenario such as that depicted in Fig.~\ref{fig:population_models}d, 
one may expect the SMC' approximation to work well when individuals are sampled from the same sub population. When the individuals are sampled from different geographical locations, however, or more generally when one may expect reduced gene flow between the different sample 
locations, the accuracy of the SMC' needs to be carefully investigated. We have found that the accuracy depends upon
the amount of gene flow within and between sub populations, on how recently divergences may have occurred, and upon
the genetic distance between the loci in question. These dependencies are summarised in Figs.~5, 6, and 7.

{\em Acknowledgments}. We acknowledge support from Vetenskapsr\aa{}det and from
the Centre for Theoretical Biology at Gothenburg University.

\bibliographystyle{elsarticle-harv}

\appendix

\section{Probability of linkage}
\label{sec:prob_linkage}

In this appendix we summarise how to calculate the probability of linkage 
in the SMC' approximation, for a freely mixing population of constant size,
and for the population-divergence model. We begin by discussing the constant population-size model, 
and then turn to the population divergence.

\subsection{Constant population size}

Consider a freely mixing population of constant size. To determine
the linkage probability of two loci in a sample of size two, the first step is to calculate the loop probability $\phi(\tau)$, the probability that 
the detached line re-attaches to the line it originated from 
before the two lines in the genealogy coalesce at time $\tau$
(upper right panel of Fig.~2). Suppose that the recombination point occurs at time $t$ (with $0 < t < \tau$). Since there are two lines to attach to, the time to attachment is exponentially distributed with rate two. Given that the line attaches before time $\tau$, it attaches to the original line with probability $1/2$. Hence, the probability of a loop occuring from recombination point at time $t$ is 
\[
\frac{1}{2} \int_t^\tau dt' 2 e^{-2(t'-t)}.
\]
In SMC' approximation, the recombination point $t$ is chosen from a uniform distribution over the gene genealogy. Hence, the probability that a recombination event re-attaches to the same line and thus leaves the gene genealogy intact (this is the loop probability) 
is found by averaging over all recombination points $t \in [0, \tau]$:
\begin{align}\label{eq:p_const}
  \phi(\tau) &=  \frac{1}{\tau} \int_0^\tau dt\, \frac{1}{2} \int_t^\tau dt' 2 e^{-2(t'-t)}
             = \frac{2\tau - 1 + e^{-2\tau}}{4\tau}\,.
\end{align}
In the second step, we use the loop probability (\ref{eq:p_const}) 
in combination with the distribution of the time to the most recent common ancestor  for a single locus to calculate the probability of linkage.
In a freely mixing population of constant size, the time to the most recent common ancestor is exponentially distributed with unit mean,
\begin{align}\label{eq:tau_distr_const}
  f(\tau) &= e^{-\tau}\,.
\end{align}
Inserting Eqs.~(\ref{eq:p_const}) and (\ref{eq:tau_distr_const}) into the general formula, Eq.~(\ref{eq:p_L_gen}), results in:
\begin{align}\label{eq:p_L_const}
  p_L(R) &= \int_0^\infty d\tau \exp\Big[\!-\tau - \left(2\tau + 1 - e^{-2\tau}\right)R/4\Big] \nonumber\\
         &= \frac{1}{2} \int_0^1 ds\,  s^{(R-2)/4} e^{-R(1-s)/4} \nonumber\\
         &= 2^{R/2} \,e^{-R/4}\, (-R)^{-(R-2)/4} \, \gamma\Big((R+2)/4,-R/4\Big)\,.  
\end{align}
Here $\gamma$ is the so-called \lq lower incomplete Gamma function'
defined as $\gamma(a,b) = \int_0^b{\rm d}t \,t^{a-1}{\rm e}^{-t}$.
Eq. (\ref{eq:p_L_const}) is our result for the probability
of linkage for two loci in a sample of two individuals
for a freely mixing population of constant size in the
SMC' approximation.

\subsection{Population divergence model}
\label{sec:Population_split_model_derivation}

For the case of a population divergence, the calculation is complicated by the difference in population structure during different stages of the history. We assume that the two individuals are sampled from different sub-populations, so that the most recent common ancestor 
occurred before the divergence. The time to the most recent common ancestor for the left-most locus has a shifted exponential distribution:
\begin{align}\label{eq:t_mrca_split}
	f(\tau) =\left\{ \begin{array}{ll}
  e^{-(\tau - D)} & \mbox{for $\tau > D$}\\
  0 & \mbox{otherwise}\,.
\end{array}\right .
\end{align}
If recombination and re-attachment occur during the divergence the result must be a loop, because the line corresponding to the other individual resides
in the other sub-population. If recombination or re-attachment do not occur during the divergence, the re-attachment process is the same as in Sec. A.1:
the line attaches with rate $2$ until the most recent common ancestor is reached.

Let $\psi(t,\tau)$ be the probability of a loop occuring, where $t$ is the time of the  recombination event and $\tau$ is the time of the most recent common ancestor. First, consider the case of $t < D$. During the interval $[t,D]$, the line re-attaches to the ancestral recombination graph with rate one, and if it re-attaches it is to the original line it detached  from. Within the time interval $[D,\tau]$, the line re-attaches with rate 2, and the probability that the re-attachment point is on the original line is $1/2$. This gives the contribution 
\begin{align}
	\psi(t,\tau) &= 1 - e^{-(D-t)} + \frac{1}{2} e^{-(D-t)} \left[ 1-e^{-2 (\tau - D) } \right] 
\end{align}
to the loop probability. When the recombination point is before the divergence ($D < t < \tau$), we obtain
\begin{align}
   \psi(t,\tau) &= \frac{1}{2} \left[1-e^{-2 (\tau -t)}\right]
\end{align}
using the same arguments as for $t < D$. Since the starting point is chosen uniformly from the gene genealogy, the loop probability $\phi(\tau)$ is
\begin{align}\label{eq:p_loop_split}
   \phi(\tau) &= \frac{1}{\tau} \int_0^\tau dt\,\psi(t,\tau) \nonumber\\
              &=\frac{2 (D+\tau) - 3 + 2 e^{-D}-\left(1-2 e^{-D}\right) e^{-2 (\tau - D) }}{4\tau}\,.
\end{align}
We insert Eqs.~(\ref{eq:t_mrca_split}) and (\ref{eq:p_loop_split}) into the general expression for the loop probability, Eq.~(\ref{eq:p_L_gen}). Further, we 
perform a change of integration variable from $\tau$ to $s = \exp[-2(\tau - D)]$. This leads to Eq.~(\ref{eq:p_L_split}).

\subsubsection{Upper bound for $p_L$}
\label{sec:pL_upper_bound}

We now show how the bound
(\ref{eq:p_L_split_approx}) was obtained.
We make
use of the fact that the exponential function is convex
and obtain the following inequality valid for $0 \le s \le 1$, with equality only at the end points:
\begin{align}
  e^{-R(1-2e^{-D})(1+s)/4} & \le e^{-R(1-2e^{-D})/4}(1-s) + e^{-R(1-2e^{-D})/2} s\,.
\end{align}
Inserting this expression into Eq.~(\ref{eq:p_L_split}) results in Eq.~(\ref{eq:p_L_split_approx}).
Plotting the right-hand side and comparing to the exact integral in Eq.~(\ref{eq:p_L_split}) shows that the inequality Eq.~(\ref{eq:p_L_split_approx}) is very close to being an equality for the parameters used in Fig.~\ref{fig:pop_split_corr}.

\subsubsection{Lower bound for $p_L$ at $D=0$}
\label{sec:pL_lower_bound}

For $D = 0$, the upper bound does not decrease exponentially with $R$, but in order to say that $p_L$ does not decrease exponentially we need a lower bound. We can obtain one such by writing (\ref{eq:p_L_const}) as
\begin{align}
	p_L = \frac{1}{2}\int_0^1 dt\, (1-t)^{(R-2)/4} e^{-Rt/4}\,.
\end{align}
Using the inequality $e^{-\alpha t} \ge (1-t)^\alpha$, which is valid for all $\alpha \ge 0$ and $0\le t \le 1$, we obtain
\begin{align}
	p_L \ge \frac{1}{2}\int_0^1 dt\, (1-t)^{(R-1)/2} =\frac{1}{1+R}
\end{align}
which we recognize as the probability of linkage in the SMC model.
From this bound it is clear that $p_L$ decreases as $1/R$ for large values of $R$, rather than decreasing exponentially
as the exact result does, compare Eqs. (\ref{eq: 3}) and ({\ref{eq:p_L_split_approx}).

\end{document}